\documentclass[numberedheadings]{aipproc}

\layoutstyle{8x11double}

\newcommand\aap{A\&A}
\newcommand\apj{ApJ}
\newcommand\apjl{ApJ}
\newcommand\pasp{PASP}

\begin{document}

\title{X-Ray Timing of the Young Pulsar in 3C~58}

\author{Scott Ransom}{ address={Department of Physics, McGill
    University, Montreal, QC H3A 2T8, Canada}, altaddress={Center for
    Space Research, Massachusetts Institute of Technology, Cambridge,
    MA 02139} }

\author{Fernando Camilo}{ address={Columbia Astrophysics Laboratory,
    Columbia University, 550 West 120th Street, New York, NY 10027} }

\author{Victoria Kaspi}{ address={Department of Physics, McGill
    University, Montreal, QC H3A 2T8, Canada}, altaddress={Center for
    Space Research, Massachusetts Institute of Technology, Cambridge,
    MA 02139} }

\author{Patrick Slane}{ address={Harvard-Smithsonian Center for
    Astrophysics, 60 Garden St, Cambridge MA, 02138} }

\author{Bryan Gaensler}{ address={Harvard-Smithsonian Center for
    Astrophysics, 60 Garden St, Cambridge MA, 02138} }

\author{Eric Gotthelf}{ address={Columbia Astrophysics Laboratory,
    Columbia University, 550 West 120th Street, New York, NY 10027} }

\author{Stephen Murray}{ address={Harvard-Smithsonian Center for
    Astrophysics, 60 Garden St, Cambridge MA, 02138} }

\begin{abstract}
  PSR~J0205$+$6449 is a young pulsar in the Crab-like pulsar wind
  nebula 3C~58 which is thought to be a result of the historical
  supernova SN1181~CE. The 65.7-ms pulsar is the second most energetic
  of the known Galactic pulsars and has been shown to be remarkably
  cool for its age, implying non-standard cooling processes in the
  neutron star core.  We report on {\em RXTE} timing observations
  taken during AO7 and supplemented by monthly radio observations of
  the pulsar made with the Green Bank Telescope (GBT).  The total
  duration covered with the timing solutions is 450\,days.  We measure
  very high levels of timing noise from the source and find evidence
  for a ``giant'' glitch of magnitude $\Delta\nu/\nu \sim
  1\times10^{-6}$ that occurred in 2002 October.  We have also
  measured the phase-resolved spectra of the pulsations and find them
  to be surprisingly hard, with photon indices $\Gamma =
  0.84^{+0.06}_{-0.15}$ for the main pulse and $\Gamma =
  1.0^{+0.4}_{-0.3}$ for the interpulse assuming an absorbed power-law
  model.
\end{abstract}

\maketitle

\section{Introduction}

PSR~J0205$+$6449 was discovered in the heart of 3C~58 in {\em
  Chandra}~HRC data and immediately confirmed in archival {\em RXTE}
data by Murray et al. \cite{mss+02}.  Camilo et al. \citep{csl+02}
soon thereafter found extremely faint radio pulsations from the pulsar
using the new Green Bank Telescope (GBT).  {\em Chandra}~ACIS-S
observations taken by Slane, Helfand, \& Murray \citep*{shm02} showed
that the temperature of the neutron star (NS) is significantly below
that predicted by standard NS cooling models given the very young age
of the pulsar ($\sim820$\,yrs).  Measurement of the spin period
(65.686\,ms) and spin-down rate ($\dot P = 1.94\times 10^{-13}s/s$) of
the pulsar have allowed the estimation of several key parameters using
the canonical expressions from pulsar theory:

\begin{itemize}
\item Spin-down energy
  \begin{equation}
    \dot{E} = 4\pi ^{2}I\dot{P}/P^{3} = 2.7\times10^{37}\,{\rm erg/s}
  \end{equation}
\item Surface magnetic field strength
  \begin{equation}
    B = \sqrt{\frac{3c^{3}IP\dot{P}}{8\pi ^{2}R^{6}}} = 3.6\times10^{12}\,{\rm Gauss}
  \end{equation}
\item Characteristic age
  \begin{equation}
    \tau = P/(2\dot{P}) = 5400\,{\rm yr}
  \end{equation}
\end{itemize}

\noindent Each of these parameters is typical for young pulsars.  One unusual
point is that if the pulsar really is 822\,yrs old and has been
spinning down with the canonical braking index (due to magnetic dipole
braking) of $n=3$, then it must have been born spinning relatively
slowly with $P_{0}\sim 60$\,ms.  Interestingly, the 65-ms
PSR~J1811$-$1925 in G11.2$-$0.5 seems to be an identical twin in many
respects (see Roberts et al., these proceedings).

\section{Observations and Data Preparation}

We analyzed all available {\em RXTE} PCA data on 3C~58, which totaled
more than $\sim$410\,ks of on-source time.  The data came from OBSIDs
P20259 ($\sim$20\,ks), P60130 ($\sim$100\,ks), and P70089
($\sim$300\,ks), and were taken in GoodXenon mode which provided
events with time resolutions of ~$\sim$122\,$\mu$s and 256 energy
channels in the energy range $\sim$2$-$60\,keV.  Most observations
were taken using 3~PCUs.

\begin{figure}[b]
  \includegraphics[height=.35\textheight,angle=270]{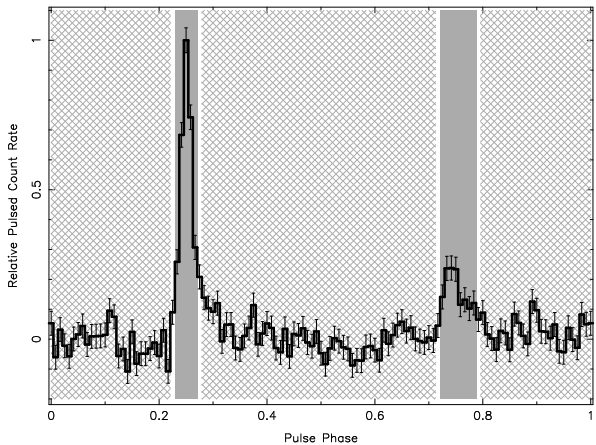}
  \caption{{\bf RXTE Pulse Profile.} The 2$-$16\,keV pulse profile for 
    PSR~J0205$+$6449 for {\em RXTE} OBSIDs P20259, P60130, and P70089.
    The total integration time included over 400\,ks with 3 PCUs and
    over 100\,ks with either 4 or 5 PCUs.  The solid grey bands show
    the portions of the pulse used for the phase-resolved spectroscopy
    described in \S\ref{sec:spect} with the pulse on the left and the
    interpulse on the right.  The lightly hashed regions were used to
    compute the spectral background.}
  \label{fig:profile}
\end{figure}

We processed the {\em RXTE} data using standard {\tt
  FTOOLs}\footnote{\url{http://heasarc.gsfc.nasa.gov/docs/software/ftools/ftools_menu.html}}
in order to filter out non-Level~1 events or those recorded outside of
Good Time Intervals and to convert their arrival times to the Solar
System barycenter (using the DE200 planetary ephemeris \citep{sta82}).
We determined the X-ray pulse times-of-arrival (TOAs) with a
maximum-likelihood technique that we developed which assumed a
two-Gaussian model of the X-ray pulse profile.  A signal-to-noise
ratio optimization procedure allowed us to determine that the best
energy range to use for the timing analysis is $\sim$2$-$16\,keV
(energy channels 4$-$39 inclusive), and so we filtered out events
outside of this energy range (see the profile in
Figure~\ref{fig:profile}).  A single time-of-arrival (TOA) typically
comprised one or two {\em RXTE} orbits depending on the background
levels and the number of PCUs on.  Since the observations consisted of
between 4$-$6 orbits, we measured 3$-$4 TOAs each month with
precisions of 100$-$200\,$\mu$s each.

The GBT data were taken with the Berkeley-Caltech Pulsar Machine
(BCPM)\footnote{\url{http://www.gb.nrao.edu/~dbacker}} backend at
either 820\,MHz or 1400\,MHz.  The BCPM is an analog/digital
filterbank which samples each of 2$\times$96 channels using 4-bits at
flexible sampling rates and channel bandwidths \citep{bdz+97}.  The
data were recorded using 134\,MHz of bandwidth and 50\,$\mu$s samples
for the 1400\,MHz observations or 50\,MHz of bandwidth and 72\,$\mu$s
samples for the 820\,MHz observations.  Typical integrations lasted
between 4$-$8\,hrs.  We de-dispersed and folded all of the data using
the pulsar analysis package {\tt PRESTO} \citep{ran01}.  Each
observation yielded 2$-$3 TOAs which we measured by correlating (in
the frequency domain) the folded pulse profile with a Gaussian of
fractional width 0.04 in phase.  The typical precision of the radio
TOAs was 200$-$400\,$\mu$s.

\section{Timing}
\label{sec:timing}

The timing behavior and therefore the spin-down rate of
PSR~J0205$+$6449 is extremely noisy.  Determining a phase-connected
timing solution would have been impossible using the {\em RXTE} or GBT
data alone.  Thankfully, the cadence of the monthly observations in
the radio and X-ray bands was sufficiently different such that the
observations complemented each other and usually allowed unambiguous
pulse numbering between observations.

We applied time corrections to transform all the TOAs to UTC at the
Solar System Barycenter using {\tt faxbary} and {\tt
  TEMPO}\footnote{\url{http://pulsar.princeton.edu/tempo}} for the
{\em RXTE} and GBT data respectively. This absolute timing shows that
the X-ray main pulse is delayed relative to the radio pulse (Camilo et
al., in prep).  We performed all timing model fits with {\tt TEMPO}.

\begin{figure}
  \includegraphics[height=.32\textheight,angle=270]{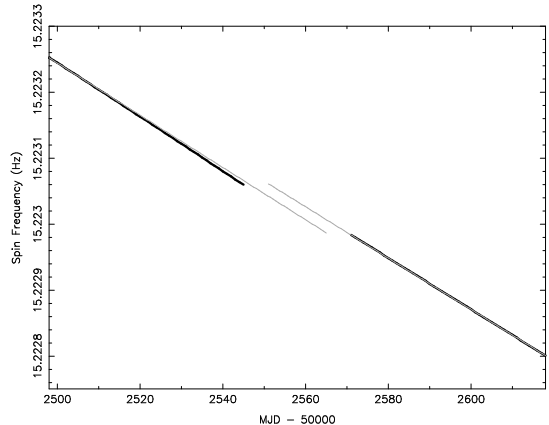}
  \caption{{\bf A ``Giant'' Glitch.} The thick lines correspond to the 
    phase-coherent timing models described in \S\ref{sec:timing}.  The
    thin grey lines correspond to the average spin-down as measured
    during the $\sim$210 days before and after MJD~52555.  The
    fractional difference in spin frequency between the two timing
    solutions is $\Delta\nu/\nu \sim $1$\times$10$^{-6}$, which is
    comparable in magnitude to the glitches from the older and also
    ``cool'' Vela pulsar.}
  \label{fig:glitch}
\end{figure}

\begin{figure}
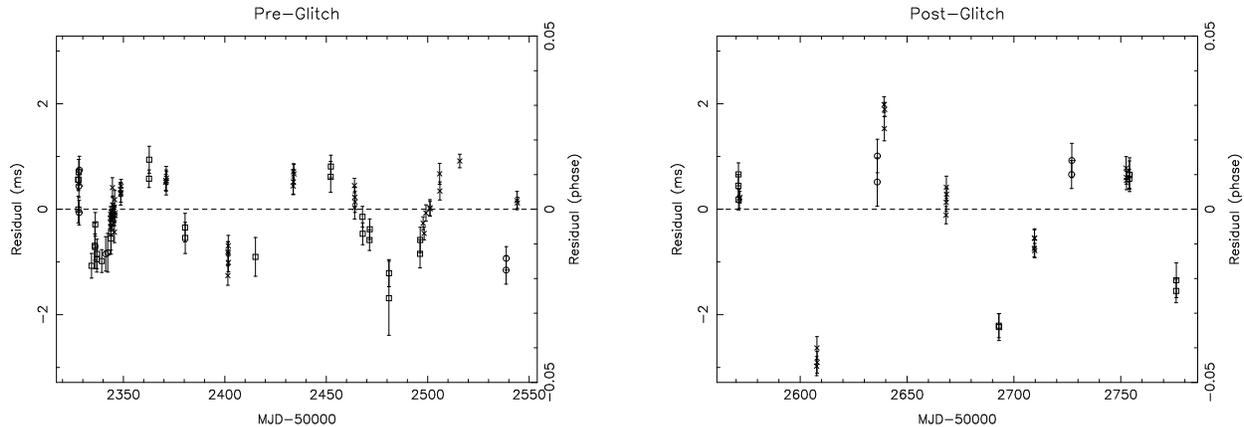

  \includegraphics[height=.34\textheight,angle=270]{ransoms_f3a}
  \hspace{1cm}
  \includegraphics[height=.34\textheight,angle=270]{ransoms_f3b}
  \caption{{\bf Post-fit Timing Residuals.} Timing residuals for pre-glitch (left) 
    and post-glitch (right) timing solutions.  TOA precision is about
    3$\times$ better with {\em RXTE} (the ``X''s) than with the GBT at
    820\,MHz (the circles) at 1400\,MHz (the squares) per unit of
    telescope time.  The obvious systematic trends are the result of
    very large levels of timing noise from the pulsar.  We lost phase
    connection near MJD~52555 due the glitch shown in
    Figure~\ref{fig:glitch}.  In both the pre- and post-glitch
    solutions, the measured braking index is much greater than the
    canonical value of 3, which is typical of timing noise or glitch
    recovery.}
  \label{fig:resids}
\end{figure}

Our data show that a ``giant'' glitch ($\Delta\nu/\nu \sim
$1$\times$10$^{-6}$) occurred near MJD~52555 (see
Figure~\ref{fig:glitch}). The timing solution prior to the glitch
consists of a 6 parameter fit (spin frequency $\nu$ plus 5 frequency
derivatives) resulting in RMS timing residuals of $\sim$0.6\,ms.  The
post-glitch timing solution consists of the spin frequency plus 2
frequency derivatives resulting in RMS timing residuals of
$\sim$1.2\,ms.  Additional frequency derivatives in the post-glitch
solution improve the residuals but are not required for phase
connection\footnote{While we believe that the post-glitch data is
  phase-connected, the sparse sampling and high levels of timing noise
  make it possible that we have miscounted pulses between one or more
  of the observations.}.  The residuals from the GBT and {\em RXTE}
for both timing solutions are shown in Figure~\ref{fig:resids}.

\section{Phase-Resolved Spectroscopy}
\label{sec:spect}

\begin{figure}
  %\hspace{0.6cm}
  \includegraphics[height=.53\textheight,angle=270]{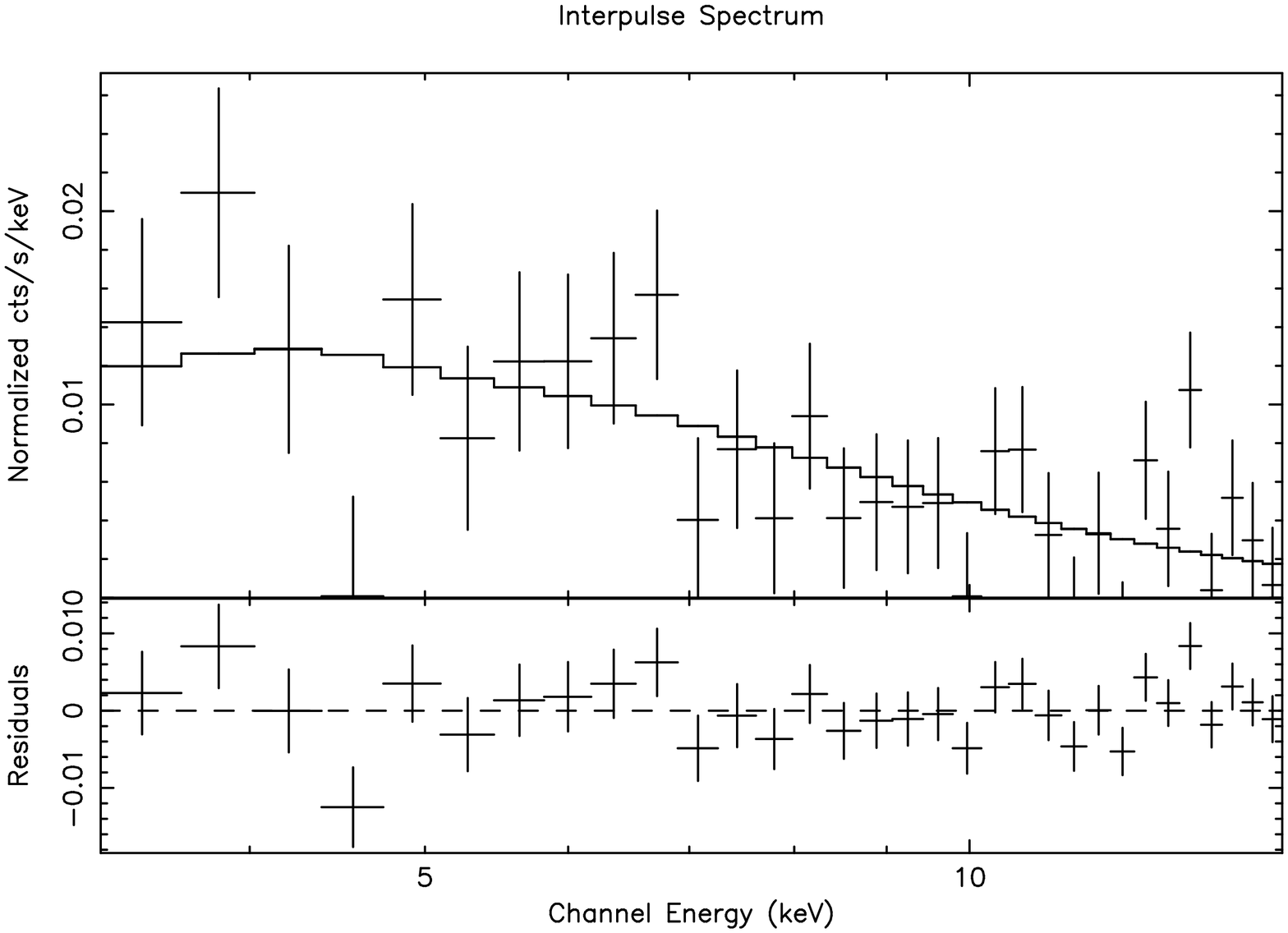}
  \caption{{\bf Phase-Resolved Spectroscopy.}  The spectra of the main X-ray 
    pulse (top) and interpulse (bottom) as fit from 3$-$16\,keV with
    an absorbed power-law model.  The $N_H$ was fixed at
    3.62$\times$10$^{21}$\,cm$^{-2}$ based on the recent 350\,ks {\em
      Chandra} ACIS observation of 3C~58 (Slane et al., in prep).  The
    best fit photon indices are surprisingly hard: $\Gamma =
    0.84^{+0.06}_{-0.15}$ for the main pulse and $\Gamma =
    1.0^{+0.4}_{-0.3}$ for the interpulse.}
  \label{fig:spect}
\end{figure}

Figure~\ref{fig:spect} shows the results of absorbed power-law fits to
the phase-resolved spectra of the pulsar's main pulse and interpulse
as determined using the {\tt FTOOL fasebin}.  We analyzed events from
each PCU independently for each OBSID, combined these for each AO, and
then combined the three AOs together.  We fit the spectra with {\tt
  xspec} after subtracting the average off-pulse background as shown
in the hatched region in Figure~\ref{fig:profile} and freezing the
value of $N_H = 3.62\times10^{21}$\,cm$^{-2}$ based on the recent {\em
  Chandra} ACIS observation of 3C~58 (Slane et al., in prep).  Fits
made to the summed PCU data and simultaneously to the 5 PCUs
independently were consistent within the statistical errors to one
another.

The measured photon indices $\Gamma = 0.84^{+0.06}_{-0.15}$ for the
main pulse and $\Gamma = 1.0^{+0.4}_{-0.3}$ for the interpulse are
surprisingly hard.  These values are difficult to reconcile with the
$\Gamma = 1.73\pm0.07$ photon index measured with {\em Chandra} for
the central point source \citep{shm02}, although significant nebular
emission within the selection region could explain some of the
spectral softening.  Recent results from the 350\,ks Chandra
observation of 3C~58 place an approximate upper limit of
$1.5\times10^{14}$\,erg\,cm$^{-2}$\,s$^{-1}$ for any hard component in
the range 0.5$-$10\,keV.  But our fits to the main pulse alone predict
fluxes in the same energy band at least one hundred times larger.  We
note that our measurements are consistent within the errors with the
phase-resolved spectral fits of Gotthelf \citep{got03} ($\Gamma =
1.11\pm0.34$ for the pulsed emission) based on the {\em RXTE} data
from P60130.

\begin{figure}
% 0.36
  \includegraphics[height=.53\textheight,angle=270]{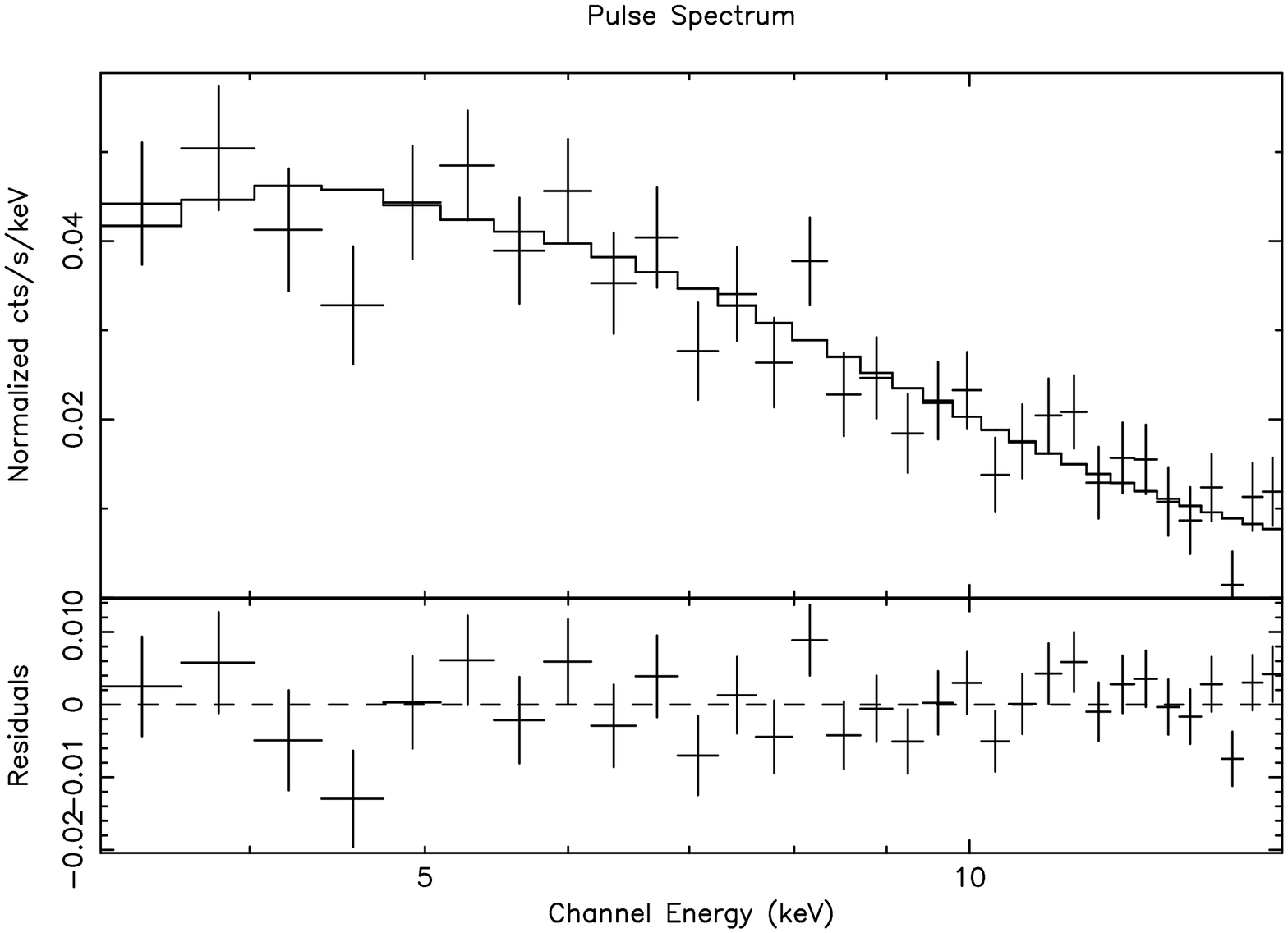}
\end{figure}

\section{Future Prospects}

With its young age, low temperature, very noisy spin-down, high $\dot
E$, hard pulsed spectrum, and large glitches, the pulsar at the center
of 3C~58 is certainly an object worth watching in the future.  Is the
extremely high level of timing noise from the pulsar somehow related
to its ``cool'' temperature?  Are both of these strange properties
related to a more fundamental characteristic of the neutron star such
as its mass?  We are optimistic that continued monitoring of this
unusual source will tell us something about the interior structure and
internal dynamics of neutron stars.  In addition, we note that the
narrow pulse profile and hard spectrum of the pulsed emission from
J0205$+$6449 make it an ideal and possibly unique target for {\em
  INTEGRAL} and other upcoming hard X-ray missions.

%\begin{theacknowledgments}
%This Thanks to all the little people.
%\end{theacknowledgments}

%\bibliographystyle{aipproc}   % if natbib is available
%\bibliography{apj-jour,pulsars,psrrefs}

\end{document}